\documentclass[10pt,journal,compsoc]{IEEEtran}

\usepackage[utf8]{inputenc}
\usepackage[T1]{fontenc}
\usepackage{amsmath,amssymb,amsfonts}
\usepackage{algorithmic}
\usepackage{graphicx}
\usepackage{textcomp}
\usepackage{soul}
\usepackage[table]{xcolor}
\usepackage{multirow}
\usepackage{balance}
\usepackage{booktabs}
\usepackage[nolist,nohyperlinks]{acronym}

\usepackage{mdframed}
\definecolor{gray90}{gray}{0.90}
\definecolor{gray95}{gray}{0.95}
\newenvironment{summary}
{\vspace{-1em}\noindent\begin{mdframed}[backgroundcolor = gray90, linecolor = lightgray, innerleftmargin=2mm, innerrightmargin=2mm]}
{\end{mdframed}}

\newcommand{\hg}{\cellcolor{green!25}}
\newcommand{\hr}{\cellcolor{red!25}}

\usepackage{subcaption}
\usepackage{url}

\usepackage[colorlinks=true,allcolors=blue]{hyperref}

\begin{acronym}
  \acro{CAIDA}{Center for Applied Internet Data Analysis}
  \acro{MAWI}{Measurement and Analysis on the WIDE Internet}
  \acro{CDN}{Content Delivery Network}
  \acro{APNIC}{Asia Pacific Network Information Centre}
  \acro{DAG}{Data Acquisition and Generation}
  \acro{WIDE}{Widely Integrated Distributed Environment}
  \acro{AS}{Autonomous System}
  \acro{MPTCP}{Multipath TCP}
\end{acronym}
\acrodefplural{CDN}[CDNs]{Content Delivery Networks}
\acrodefplural{AS}[ASes]{Autonomous Systems}

\begin{document}
\bstctlcite{IEEEexample:BSTcontrol}

\title{Impact of Evolving Protocols and\\COVID-19 on Internet Traffic Shares}

 \author{
 \IEEEauthorblockN{Luca Schumann\IEEEauthorrefmark{1}, Trinh Viet Doan\IEEEauthorrefmark{1}, Tanya Shreedhar\IEEEauthorrefmark{2}, Ricky Mok\IEEEauthorrefmark{3}, and Vaibhav Bajpai\IEEEauthorrefmark{1}}
\\ 
 \IEEEauthorblockA{\IEEEauthorrefmark{1}TUM, Germany
 \hspace{0.06in}
 \IEEEauthorrefmark{2}IIIT-Delhi, India
 \hspace{0.06in}
 \IEEEauthorrefmark{3}CAIDA, USA
 }
 }

\maketitle

\begin{abstract}
  The rapid deployment of new Internet protocols over the last few years and the
COVID-19 pandemic more recently (2020)
has resulted in a change in the Internet traffic composition. Consequently, an
updated microscopic view of traffic shares is needed to understand how the
Internet is evolving to capture both such shorter- and longer-term events. 
Toward this end, we observe traffic composition at a research network in Japan
and a Tier-1 ISP in the USA. We analyze the traffic traces passively captured at
two inter-domain links: MAWI (Japan) and CAIDA (New York--S\~{a}o Paulo), which
cover $\approx$100~GB of data for MAWI traces and $\approx$4~TB of data for
CAIDA traces in total.
We begin by studying the impact of COVID-19 on the monitored MAWI link: We find
a substantial increase in the traffic volume of OpenVPN and \texttt{rsync}, as
well as increases in traffic volume from cloud storage and video conferencing
services, which shows that clients shift to remote work during the pandemic.
For traffic traces between March 2018 and December 2018, we find that the use of
IPv6 is increasing quickly on the CAIDA monitor: The IPv6 traffic volume
increases from 1.1\% in March 2018 to 6.1\% in December 2018, while the IPv6
traffic share remains stable in the MAWI dataset at around 9\% of the traffic
volume.
Among other protocols at the application layer,
60\%--70\% of IPv4 traffic on the CAIDA link is HTTP(S) traffic, out of which
two-thirds are encrypted; for the MAWI link, more than 90\% of the traffic is
Web, of which nearly 75\% is encrypted. Compared to previous studies, this
depicts a larger increase in encrypted Web traffic of up to a 3-to-1 ratio of
HTTPS to HTTP. As such, our observations in this study further reconfirm that
traffic shares change with time and can vary greatly depending on the vantage
point studied despite the use of the same generalized methodology and analyses,
which can also be applied to other traffic monitoring datasets.

\end{abstract}

\section{Introduction}\label{sec:introduction}

Technologies and services on the Internet (and their underlying building blocks)
have changed drastically over the last decade. While many of the Internet
protocols nowadays, such as IPv6 or HTTPS, had been proposed more than 15 years
ago, their adoption has only recently \emph{evolved} and gained
traction. With more recent protocols such as \ac{MPTCP},
QUIC~\cite{DBLP:conf/sigcomm/LangleyRWVKZYKS17, kosek2021beyond}, or FB-Zero~\cite{fb-zero}
joining the mix, today's traffic composition is much different from what it was
a few years ago. These evolving protocols are essential in shaping and advancing
the Internet: IPv6 enables an increasing number of connected users, MPTCP
supports seamless mobility, and QUIC and FB-Zero drive low-latency interaction,
along with securing Web traffic together with HTTPS.
As a result, these protocols have seen increasing adoption: Google measures
25--30\% of its users accessing their servers via IPv6~\cite{google_ipv6},
although its growth has slowed down since 2016~\cite{JiaLHEACD19}. HTTPS support
has increased significantly in the last few years, with 40\% of the Alexa Top 1M
websites offering HTTPS as of 2016~\cite{DBLP:conf/uss/FeltBKPBT17}. A
study~\cite{DBLP:conf/conext/TrevisanGDMM18} from 2017 measures traffic from an
Italian ISP network and found that both QUIC and FB-Zero each carry around 10\%
of the total Web traffic. 

While these evolutions have significantly transformed the Internet traffic
composition, especially over the last decade, more short-term events such as the
recent COVID-19 pandemic have also affected the application mix, shifting
traffic toward protocols and services which are related to remote work.

To this end, we first consider related studies
(\textsection\ref{sub:rel-work-covid}) on the impact of COVID-19 on network
traffic, finding that these studies are biased regarding the observed network in
terms of types and regions. Further, previous work has not focused on changes in
source and destination \acp{AS} of traffic during the pandemic. As such, we
analyze datasets from a Japanese Internet backbone, operated by the \acf{MAWI}
Working Group~\cite{MAWI,ChoMK00} to study the impact of the COVID-19 lockdown
on the MAWI traffic mix. We compare traffic from 2019 to 2020
(\textsection\ref{sec:covid}) in order to investigate the following research
questions: \emph{How do restrictions for spread prevention affect the Internet
traffic composition? Do popular source and destination ASes of traffic change
during the pandemic, and if so, how?}

Besides such sudden changes, the Internet has also become increasingly flat over
the years, which results in most traffic only traveling one inter-AS
link~\cite{DBLP:conf/imc/ChiuSRKG15,ArnoldHJCCGK20}. Thus, it is also necessary
to revisit and re-quantify the composition of Internet traffic
(\textsection\ref{sub:rel-work-mix}) in a broader scope. However, as
aforementioned studies have shown, observations can differ quite significantly
based on the location of the vantage
point~\cite{google_ipv6-bycountry,apnic,DBLP:conf/uss/FeltBKPBT17}, and are thus
not necessarily generalizable. Further, traffic over inter-domain links that
capture such evolutions has not been extensively studied yet, although it is
important to understand the usage (or lack thereof) of such evolving protocols
for traffic engineering purposes. We focus on these evolving protocols, as they
represent pressing issues of this decade, which is why trends of their evolution
are of particular interest for the future of the Internet. Consequently, we pose
the following research questions with respect to such links while contrasting
two different vantage points: \emph{What is the traffic share of IPv6? Has HTTPS
already replaced HTTP? Have the newer transport layer protocols replaced
traditional protocols and to what extent? What other applications contribute
significantly to the application mix?}

To answer these questions, we leverage monitoring data from an inter-domain link
between New York and S\~{a}o Paulo, collected by the
\acf{CAIDA}~\cite{CAIDA:monitor,CAIDA:dataset} to focus on such evolving
protocols, and contrast the results with traces collected in the same time
period from the MAWI monitor.

The MAWI backbone carries a mix of traditional and experimental research
traffic; the CAIDA monitor captures traffic from a Tier-1 ISP backbone link in
both directions and, thus, carries more conventional traffic
(\textsection\ref{sec:dataset}). Analyzing the traffic monitored on these links
allows us to investigate the usage and impact of evolving protocols on the
Internet traffic composition, while also highlighting that trends differ based
on the observed link, and even on the observed link
direction~(\textsection\ref{sec:longitudinal}).
In our analysis, we focus on
byte shares to denote the traffic percentages. We then discuss
limitations~({\textsection\ref{sec:limitations}}) before concluding the
study~({\textsection\ref{sec:conclusion}}). Our primary findings are:

\noindent\textbf{--- Impact of COVID-19:} Comparing the MAWI traffic mix from
April 2019 to the one from April 2020, we find that although the overall number
of flows has increased, the number of bytes transmitted over the research
network link has reduced by 47.3\% over IPv4 and 38.5\% over IPv6. Further,
differences in traffic volume between weekdays and weekend have become less
pronounced in 2020. For the heavy-hitting applications, we observe that the
volume differences of protocols related to remote access
(\textsection\ref{sub:covid-mix}) such as OpenVPN and \texttt{rsync} increase
significantly over both address families when comparing 2019 and 2020 (by up to
3082.7\%), although \texttt{ssh} and X11 exhibit lower traffic volume in 2020
(lower by 63.7--84.0\%). 
In terms of source ASes for both 2019 and 2020
(\textsection\ref{sub:covid-ases}), we find that traffic is primarily received
from from \acp{CDN} and cloud providers, although the volume drastically
decreases from 2019 to 2020. However, traffic volume incoming from Dropbox
(AS19679) increases in 2020, along with increases of traffic sent toward Cisco
Webex (AS13445), which indicate a shift to remote work among users.

\noindent\textbf{--- Impact of Evolving Protocols:}
On the CAIDA link, IPv6 accounts for 4\% of the traffic (byte share) on average
in both directions, although the IPv6 traffic shares are highly asymmetric when
comparing the individual directions of the link. In comparison, the MAWI link
shows a significantly higher IPv6 traffic share (9\% average in 2018, however,
increasing to 12.2\% in 2020) (\textsection\ref{sub:analysis-IP}). Further, the
MAWI IPv6 mix converges to its IPv4 mix, which indicates that both address
families are used similarly.

TCP is still the dominant transport protocol (91.5\% of traffic share) on the
monitored links. Although, TCP usage is declining at the expense of evolving
transport protocols, such as FB-Zero and QUIC, since QUIC already achieves 7\%
traffic share. Further, we only find 1.5\% of the traffic being carried over UDP
(besides QUIC), indicative of its usage primarily for carrying DNS traffic or as
a multiplexer to support the deployment of new transport protocols on top of its
stack (\textsection\ref{sec:transport_prot}).

We notice a growth of HTTPS traffic, while HTTP traffic declines. Thus, the
ratio of HTTPS to HTTP increases, reaching ratios of 2-to-1 up to 3-to-1 in
favor of encrypted Web traffic. As QUIC and FB-Zero primarily carry encrypted
Web traffic and simultaneously experience higher usage, the traffic share of
HTTPS increases consequently (\textsection\ref{sec:Web_dominance}).

\bigskip\noindent
Overall, our study is directly relevant for network and operational communities,
since knowledge of both short-term and long-term trends in Internet traffic
composition is crucial not only for better management of networks and traffic
engineering, but also for predicting future trends to allow ISP networks and
content providers to be better prepared either via reconfiguration or via
increasing network capacity. The respective contributions and novelty of the
paper are three-fold:

\begin{enumerate}
    \item The first contribution is the evaluation of how traffic shares change
    on the Internet due to both short-term (COVID-19 lockdown measures) and
    long-term (evolving protocols) effects. Previous work has majorly evaluated
    either effect, however, not both in the same study. We show that studies
    need to observe both effects together, to be able to provide a more
    comprehensive view on Internet traffic shares. 
    \item Second, analyses should consider different vantage points on the
    globe. For this, we compare the Internet traffic shares at two distinct
    vantage points (West and East) on the globe during the same time period,
    after which we show that traffic shares vary significantly based on
    location, which also applies to the evolving protocols in these regions.
    Previous work has majorly focused on traffic shares by analyzing dataset of
    a particular location or an AS.
    \item Lastly, despite observing different trends for the two different
    traces, we demonstrate that our method and analyses are generalized and can
    be applied to other traffic monitoring datasets. This is a valuable
    contribution to the community, since the method and analyses are not
    tailored to a specific dataset.
\end{enumerate}

\section{Related Work}\label{sec:related_work}

\subsection{Effects of the COVID-19 on Internet Traffic}
\label{sub:rel-work-covid}

After the outbreak of the COVID-19 pandemic, several studies have investigated
the impact of countermeasures to the pandemic on the Internet traffic.

Favale \emph{et al.}~\cite{FavaleSTDM20} analyze the effect on the campus
network of an Italian university, which implemented e-learning solutions for
classes. They specifically focus on passive traces and application logs of
collaborative tools such as Microsoft Teams and tools for remote access such VPN
and remote desktop services. They find that the amount of incoming traffic has
decreased significantly, whereas the outgoing traffic (to the student's homes)
has increased by more than a factor of two.

A similar picture is shown by B\"ottger \emph{et al.}~\cite{Boettger2020}, who
use Facebook's edge network around the globe to reveal that different regions
have been affected differently by the lockdown measures. Similarly, the traffic
demand in terms of volume increased, whereas the traffic demand in terms of
application type also shifted toward live streaming and messaging, both
following changes in user behavior at the edge.

In contrast, Lutu \emph{et al.}~\cite{Lutu2020} study the effect of the COVID-19
pandemic on a mobile network in the UK. Due to the limited mobility of users as
a result of lockdown measures, they identify that both mobility and download
traffic volume decreased, although these changes also differ between
geographical locations and social backgrounds of users. Moreover, they see an
increase in voice traffic, which overall indicates that the types of used
services have shifted.

Feldmann \emph{et al.}~\cite{Feldmann2020} leverage multiple datasets from a set
of IXPs, an ISP, an an educational network to study implications of the outbreak
on the Internet traffic. In general, they find an increase in traffic volume of
15--20\% in a week, as well as traffic shifts away from hypergiants and toward
applications typically used at home and for remote work (like VPNs or Web
conferencing).

Gives these different observations from different networks around the world,
e.g., in Italy, where the initial outbreak was more severe, an additional study
regarding the impact on the Japanese research networks monitored by MAWI
provides a complementary view for the overall picture of the pandemic's effect
on Internet traffic. However, note that observations are not directly comparable
due to inherent differences of the monitored networks.

\subsection{Traffic Share of Protocols}
\label{sub:rel-work-mix}
Felt \emph{et al.}~\cite{DBLP:conf/uss/FeltBKPBT17} analyzed HTTPS and HTTP
measurements between 2014 and 2017 from different viewpoints, one of which was
the MAWI monitor that we also use in this work. They saw an ongoing increase in
HTTPS traffic and a decrease in HTTP traffic over the three years, similar to
previous measurements~\cite{DBLP:conf/pam/RichterCSFW15}; for example, the
percentage of HTTPS traffic grew from 20\% to 40\% of Web traffic. Chan \emph{et
al.}~\cite{DBLP:conf/infocom/ChanFCG18} present similar results for their
analysis of the same traces between 2009 and 2017. Further, they classified the
HTTPS traffic per major AS in the time span, where Amazon showed to be the main
HTTPS contributor with Facebook right after.

Transport protocols have seen multiple advancements in the last
decade~\cite{PoleseCBRZZ19}. As for
QUIC~\cite{DBLP:conf/sigcomm/LangleyRWVKZYKS17}, previous measurement
studies~\cite{DBLP:conf/icc/MegyesiKM16,DBLP:conf/globecom/BiswalG16,DBLP:conf/icc/CookMTH17,DBLP:conf/ipccc/YuXY17,shreedharEvaluating}
primarily focused on the performance aspects, with only a few studies looking
into measuring the usage and adoption of QUIC. R\"uth \emph{et
al.}~\cite{RuthPDH18} studied the usage of QUIC within the IPv4 address space,
finding traffic shares of 2.6\%--9.1\% for different vantage points. They
further showed that Google is the main driver of QUIC. Trevisan \emph{et
al.}~\cite{DBLP:conf/conext/TrevisanGDMM18} used data from an Italian ISP
gathered between 2012 and 2017 to give an overview of the Internet evolution.
They noticed that Internet giants such as Google and Facebook could quickly deploy
new transport protocols (namely QUIC~\cite{DBLP:conf/sigcomm/LangleyRWVKZYKS17}
and FB-Zero~\cite{fb-zero}), leading to them together carrying around 20-25\% of
Web traffic at the end of the measurement period. Regarding MPTCP, a recent study~\cite{aschenbrennersingle} reveals a steady increase in MPTCP-capable IPs. While the MPTCP usage has increased, its traffic share remains fairly low.


There are varying reports regarding the IPv6 adoption on the Internet, as the
use of different metrics leads to different
observations~\cite{DBLP:conf/sigcomm/CzyzAZIOB14}. Akamai reported 11\% IPv6
traffic in July 2017~\cite{akamai_internet_report}. Google measured 19\% IPv6
connectivity on their clients in July 2017 and around 30\% in January
2021~\cite{google_ipv6}. They also detect a 44.4\% IPv6 capability in the USA
and 36.8\% in Brazil for January 2021. At the same time, the \ac{APNIC} measures
a 53.7\% (USA) and 36.7\% (Brazil) IPv6 capability, respectively~\cite{apnic}.
Although IPv6 has exponentially grown since
2012~\cite{DBLP:conf/sigcomm/CzyzAZIOB14,DBLP:conf/pam/RichterCSFW15}, the
growth has slowed down substantially since 2016~\cite{JiaLHEACD19}. On the other
hand, reports based on global corporate data~\cite{Sandvine:report,Cisco:report}
primarily focus on other metrics, such as number of connected users or
performance, and application usage, rather than traffic shares of underlying
protocols. Moreover, note that protocol support is different from protocol
usage.

Considering the varying numbers observed by different studies, it is important
revisit the traffic composition while considering differences due to the
selection of vantage points. Thus, we attempt to highlight this by contrasting
two datasets of different traffic monitors in this study. We further dissect the
data of one monitored link by its two directions to show similarities and
differences within a single link, since Internet traffic can be asymmetric as
well.

\section{Datasets and Methodology}\label{sec:dataset}

For the analysis of short-term changes, we first discuss the impact of the
restrictions imposed by the current COVID-19 pandemic on the traffic composition
monitored by MAWI~(\textsection\ref{sec:covid}), for which we compare datasets
collected by the MAWI monitor from April 2019 and April 2020.

We then analyze two sets of publicly available traffic traces to study the
natural evolution of the composition of Internet traffic at two different
geolocations. To directly contrast the two datasets with each other, we limit
the study period to monthly traces collected between March 2018 and December
2018 to align the timeline of both datasets:
For comparison of the traffic seen at the monitors
(\textsection\ref{sec:longitudinal}), we consider traces for the days on which
both CAIDA and MAWI monitors have collected data, i.e., within the same
timeframe. Note that the CAIDA monitors stopped capturing data after January
2019, as the link upgraded to 100~Gbps, which the monitoring cards are not
capable of handling anymore, which is why the common time period of the data
is limited to 2018.

Moreover, parts of the MAWI datasets from April 2019 and 2020 are not anonymized
(see~\textsection\ref{sub:covid-ases}), in contrast to the public datasets from
2018 in which source and destination endpoints are anonymized. Thus, analyses in
\textsection\ref{sec:covid} and \textsection\ref{sec:longitudinal} are
inherently different. Nevertheless, despite the data being available in the
public domain, the observable trends derived from the data analysis are not
necessarily known, which represents the novel contribution of this study.

\textbf{MAWI --} The \acf{MAWI} Working Group monitors network traffic as part
of the \acf{WIDE} project. The \ac{WIDE} project~\cite{WIDE} runs an Internet
backbone in Japan, which is an operational network and simultaneously serves as
an experimentation ground for new applications. We download the collected traces
for \texttt{samplepoint-F} ($\approx100$~GB in size for March to December 2018),
which is a 1~Gbps transit link of WIDE to the upstream ISP. The traffic traces
are collected via \texttt{tcpdump}, with confidential information being removed
via \texttt{TCPdPriv}~\cite{MAWI}. Since 2007, MAWI publishes 15-minute captures
(14:00--14:15 local time) of this link every day. They release compressed
\texttt{pcap} files for each trace, along with a summary of the packets and
bytes observed per IP protocol.

\textbf{CAIDA --}
The dataset collected by \ac{CAIDA} consists of anonymized Internet
traces~\cite{FragkouliAF19} captured from a passive monitor~\cite{CAIDA:monitor}
located in an Equinix data center in New York City. The monitor connects to a
9953 Mbps Tier-1 ISP backbone link between New York and S\~{a}o Paulo. The
packets are captured by two machines, each of them using a Endace 9.2 DAG
network monitoring card to capture one direction of the full-duplex fiber optics
link:
direction ``A'' from S\~{a}o Paulo to New York and direction ``B'' from New York
to S\~{a}o Paulo, which we refer to as \emph{SPNY} and \emph{NYSP},
respectively, for the remainder of this work, in which we present separate
observations for each direction.
After collecting the data, CAIDA strips the payload from the packets and
anonymizes the IP addresses through CryptoPan prefix-preserving
anonymization~\cite{cryptopan}. For IPv4, all 32 bits are preserved, while for
IPv6 addresses, only 64 bits are preserved.
The traces considered in this study were recorded between March 15, 2018, and
December 20, 2018 ($\approx4$~TB in size); the monitor records the traces on the
third Thursday of each month for both directions at 13:00 UTC for 1 hour. The
corresponding local time of the traces is 08:00--09:00 AM in New York and
10:00--11:00 AM in S\~{a}o Paulo. 
The recorded \texttt{pcap} files contain the anonymized traces with header
information up to the transport layer (L4) and timestamps truncated to
microsecond precision.

\textbf{Methodology --} To investigate the distribution of IP versions, we
extract the relevant meta information provided by both datasets, which includes
the bytes observed per IP version. To obtain the transport layer information,
namely the transport protocol (such as TCP/MPTCP/UDP/QUIC), port numbers, and IP
addresses (source and destination)
of the packets, we analyze the \texttt{pcap} files using the \texttt{SiLK}
analysis suit~\cite{netsa} and \texttt{TShark} network protocol
analyzer~\cite{tshark}.
For both MAWI and CAIDA traces, a small fraction (around 0.5\%) of packets are lost during the
conversion due to incomplete packet headers. Note that the CAIDA data collected in October 2018 has
around 50\% incomplete headers; for those, we manually extracted the \texttt{pcap} files of this month
instead.

Throughout our results, we refer to the number of bytes transferred when we mention quantities of
traffic. To extract the information on applications in the dataset, we apply a heuristic
that maps tuples of transport protocol and port to known applications (e.g., UDP and port 53 are mapped to DNS).
We first extract the top 50 combinations of transport protocol with source and
destination ports from all traces for each month and count the monthly
occurrence of every combination.
Note that for the CAIDA dataset, we consider and process the two directions SPNY
and NYSP separately. For the top 50 combinations, we further normalize the data
between the CAIDA and MAWI data and also discard applications that have only
become popular over a brief period of time.
Note that not all combinations can be mapped to one specific application; e.g.,
some combinations can be mapped to several applications, while others cannot be
mapped to any application. Many applications often use a wide range of ports
(several hundred), making it difficult to classify with anonymized headers. We
thus map the port/protocol combinations on a best effort basis.

\section{Impact of COVID-19}
\label{sec:covid}
\begin{table*}[!t]
\caption{\emph{Aggregates of MAWI samplepoint-F traffic volume in GB over IPv4
(left) and IPv6 (right) for a subset of applications as of April 2019 and April
2020, showing relative volume difference in \% and share difference in
percentage points (p.p.).}
}
\label{tab:apr-19vs20}
\centering
\resizebox{0.8\textwidth}{!}{%
\begin{tabular}{lrrrrrr}
    \toprule
    \multicolumn{1}{c}{\textbf{\begin{tabular}[c]{@{}c@{}}IPv4\\ Application\end{tabular}}}
    & \multicolumn{1}{c}{\textbf{\begin{tabular}[c]{@{}c@{}}Traffic Vol.\\ (2019) {[}GB{]}\end{tabular}}}
    & \multicolumn{1}{c}{\textbf{\begin{tabular}[c]{@{}c@{}}Share\\ (2019)\end{tabular}}}
    & \multicolumn{1}{c}{\textbf{\begin{tabular}[c]{@{}c@{}}Traffic Vol.\\ (2020) {[}GB{]}\end{tabular}}}
    & \multicolumn{1}{c}{\textbf{\begin{tabular}[c]{@{}c@{}}Share\\ (2020)\end{tabular}}}
    & \multicolumn{1}{c}{\textbf{\begin{tabular}[c]{@{}c@{}}Relative\\ Vol. Diff.\end{tabular}}}
    & \multicolumn{1}{c}{\textbf{\begin{tabular}[c]{@{}c@{}}Share\\ Diff.\end{tabular}}}
    \\ \midrule
                   HTTPS &    1077.568 &     54.2\% &     476.971 &     45.5\%    & $-$55.7\%\hr  &  \hr$-$8.6 p.p.  \\
                    HTTP &     502.436 &     25.3\% &     292.597 &     27.9\%    & $-$41.8\%\hr  &  \hg$+$2.7 p.p.  \\
                     SSH &      35.612 &      1.8\% &       5.687 &      0.5\%    & $-$84.0\%\hr  &     $-$1.2 p.p.  \\
         IPSec NAT Trav. &       4.194 &      0.2\% &       5.537 &      0.5\%    & $+$32.0\%     &     $+$0.3 p.p.  \\
                     DNS &       2.653 &      0.1\% &       2.433 &      0.2\%    & $-$8.3\%      &     $+$0.1 p.p.  \\
                   rsync &       2.328 &      0.1\% &       5.352 &      0.5\%    & $+$129.9\% \hg&     $+$0.4 p.p.  \\
                 OpenVPN &       0.664 &      0.0\% &       7.125 &      0.7\%    & $+$973.4\% \hg&     $+$0.6 p.p.  \\
                     X11 &       0.635 &      0.0\% &       0.230 &      0.0\%    & $-$63.7\% \hr &     $-$0.0 p.p.  \\
                     FTP &       0.094 &      0.0\% &       0.115 &      0.0\%    & $+$22.6\%     &     $+$0.0 p.p.  \\  \midrule
          \textbf{Total} &    1989.694 &      100\% &    1048.084 &      100\%    & $-$47.3\%     &           {}  \\
    \bottomrule
    \end{tabular}
}

\vspace{1em}

\resizebox{0.8\textwidth}{!}{%
\begin{tabular}{lrrrrrr}
    \toprule
    \multicolumn{1}{c}{\textbf{\begin{tabular}[c]{@{}c@{}}IPv6\\ Application\end{tabular}}}
    & \multicolumn{1}{c}{\textbf{\begin{tabular}[c]{@{}c@{}}Traffic Vol.\\ (2019) {[}GB{]}\end{tabular}}}
    & \multicolumn{1}{c}{\textbf{\begin{tabular}[c]{@{}c@{}}Share\\ (2019)\end{tabular}}}
    & \multicolumn{1}{c}{\textbf{\begin{tabular}[c]{@{}c@{}}Traffic Vol.\\ (2020) {[}GB{]}\end{tabular}}}
    & \multicolumn{1}{c}{\textbf{\begin{tabular}[c]{@{}c@{}}Share\\ (2020)\end{tabular}}}
    & \multicolumn{1}{c}{\textbf{\begin{tabular}[c]{@{}c@{}}Relative\\ Vol. Diff.\end{tabular}}}
    & \multicolumn{1}{c}{\textbf{\begin{tabular}[c]{@{}c@{}}Share\\ Diff.\end{tabular}}} 
    \\ \midrule
                   HTTPS &     171.205 &     72.1\% &      22.887 &     15.7\%    & $-$86.6\%\hr  &   \hr$-$56.4 p.p  \\
                    HTTP &      39.365 &     16.6\% &      61.976 &     42.5\%    & $+$57.4\%\hg  &   \hg$+$25.9 p.p  \\
                     SSH &      10.492 &      4.4\% &       3.331 &      2.3\%    & $-$68.3\%     &       $-$2.1 p.p  \\
                     DNS &       1.725 &      0.7\% &       4.345 &      3.0\%    & $+$151.9\%    &       $+$2.3 p.p  \\
                   rsync &       0.478 &      0.2\% &      15.210 &     10.4\%    & $+$3082.7\%\hg&   \hg$+$10.2 p.p  \\
                     X11 &       0.022 &      0.0\% &       0.004 &      0.0\%    & $-$82.0\%     &       $-$0.0 p.p  \\
                 OpenVPN &       0.009 &      0.0\% &       0.030 &      0.0\%    & $+$219.6\%\hg &       $+$0.0 p.p  \\
                     FTP &       0.001 &      0.0\% &       0.004 &      0.0\%    & $+$249.9\%\hg &       $+$0.0 p.p  \\
         IPSec NAT Trav. &       0.000 &      0.0\% &       0.000 &      0.0\%    & $+$126.9\%\hg &       $+$0.0 p.p  \\  \midrule
          \textbf{Total} &     237.376 &      100\% &     145.944 &      100\%    & $-$38.5\%     &             {}  \\
    \bottomrule
    \end{tabular}
}
\end{table*}

Before considering ``natural'' evolutions of the application mix, we first
investigate sudden changes to the traffic composition: The COVID-19 pandemic has
resulted in a drastic shift in Internet traffic from work environments to home
networks. Governments and various institutions are recommending people to work
from home to limit the spread of the virus.
Recent studies~\cite{FavaleSTDM20,Lutu2020,Feldmann2020,Boettger2020} investigate how
COVID-19 has impacted the Internet and traffic from different networks and
vantage points, primarily from Europe. In this section, we investigate the
change in traffic composition on the MAWI link (samplepoint-F) in April 2019 and
April 2020; recall that this link carries a mix of traditional and experimental
traffic from research networks in Japan. Furthermore, Japan did not have a
lockdown for the general public (other than most countries where the vantage
points of the aforementioned studies were located). Instead, Japan has put
preventive measures into place starting April 2020; on-campus operations have
been shut down at major universities in Tokyo~\cite{keio1,titech1,utokyo1}, from
whose research network the MAWI data is collected. Thus, for countries which
implemented more strict and limiting regulations, the observed impact of the
COVID-19 pandemic on the Internet traffic are likely more extreme
(cf.~\cite{Feldmann2020,Lutu2020}). As the CAIDA link stopped monitoring data
after January 2019, traffic data collected from the link between New York and
S\~{a}o Paulo cannot be included in this analysis.

To this end, we aggregate the traffic traces across all of April for 2019 and
2020 (referred to as 2019 and 2020 in the following). We find 838.9M flows
overall for 2019, whereas for 2020, we observe 1.26B flows in total, i.e., an
increase by roughly 50\%. On the other hand, the number of packets has remained
similar with 3.1B in 2019 and 2.8B in 2020. 
Nevertheless, we observe packet sizes (i.e., bytes per packet) to be smaller, as
the total number of bytes has reduced drastically. Moreover, the common
weekday-weekend patterns with respect to traffic volumes become less pronounced
in 2020 (cf.~\cite{Feldmann2020}). The aggregates in terms of traffic volume in
bytes are shown in Table~\ref{tab:apr-19vs20} over IPv4 (left) and IPv6 (right),
along with the relative changes to the previous year (highlighted cells show
substantial changes from 2019 to 2020). Overall, we see traffic volume has
dropped significantly by 47.3\% over IPv4 and 38.5\% over IPv6 when comparing
2019 to 2020, which is expected, as traffic from the WIDE network has likely
moved to residential networks due to home office regulations. The table further
indicates that IPv6 traffic share was 10.7\% in 2019 and has increased to 12.2\%
in 2020, suggesting continuous growth in IPv6 adoption since 2018 (when the
share was around 9\%).

\subsection{Application Mix and Remote Work}
\label{sub:covid-mix}

We focus on a subset of applications that we have seen to be prominent in
previous sections or which are related to remote access. In particular, the
volume of Web traffic (HTTPS and HTTP) over IPv4 decreases by 55.7\% and 41.8\%,
respectively, although the difference in traffic share only changes
substantially for HTTPS ($-$8.6 percentage points (p.p.)). Traffic volume
attributed to \texttt{ssh} drops by 84.0\%, which also affects X11 ($-$63.7\%),
although both protocols are typically used for remote work. Nevertheless, we
observe that the traffic volume of \texttt{rsync} and OpenVPN has increased
manifold by 129.9\% and 973.4\%; especially the high increase in OpenVPN traffic
volume indicates that clients likely connect to the WIDE network through a VPN,
which then carries other applications.

Regarding IPv6, we find that HTTPS traffic volume decreases massively by 86.6\%,
resulting in a traffic share difference of $-$56.4 p.p.. In contrast, however,
HTTP traffic increases by 57.4\%, increasing traffic share by 25.9 p.p..

We suspect that the decrease of HTTPS traffic volume in the network is likely
linked to the decrease of Web service usage (e.g., YouTube, Facebook, and
Netflix that are known to be drivers of IPv6 traffic), which has reduced
together with the number of users (see~\textsection\ref{sub:covid-ases}). On the
other hand, similar to IPv4, we observe the largest changes for \texttt{rsync}
($+$3082.7\% traffic volume), FTP ($+$249.9\%), OpenVPN ($+$219.6\%), and IPSec
NAT Traversal ($+$126.9\%), with the traffic share of \texttt{rsync} even
increasing by 10.2 p.p. as a result.

Overall, we find that the traffic composition over IPv4 remains roughly the
same during the prevention measures. However, over IPv6, the composition and
rankings change, as HTTP and HTTPS switch places in terms of the highest traffic
shares. Moreover, \texttt{rsync} contributes more than 10.4\% of the traffic
over IPv6 in 2020, whereas in 2019, it only amounted to a percentage of 0.2\%,
showing a significant increase.

\subsection{Source and Destination ASes}
\label{sub:covid-ases}

We are further given access to non-anonymized traffic trace headers from MAWI
for April 2019 and 2020, which allows us to identify the traces' source and
destination \acp{AS} from the real IP addresses for these months specifically.
To map the ASes, we use BGP Routing Information Base data collected by
RouteViews for the respective months. On top of that, we lookup the types of the
ASes based on the CAIDA AS classification dataset~\cite{caida-as-types}, which
differentiates between \emph{transit/access} (T/A), \emph{content} (C), and
\emph{enterprise} (E) ASes in particular. However, note that the non-anonymized
dataset only covers Wednesdays and Sundays of the months and is therefore a
subset of the previously discussed traffic traces. We aggregate the data for
these days across the whole month and group bytes by \ac{AS}. In 2019, this
subset covers 551.7~GiB of traffic,
whereas in 2020, the subset covers only 292.4~GiB of traffic
instead. We inspect the top 10 source (i.e., incoming traffic) and top 10
destination (i.e., outgoing traffic) ASes by traffic volume, shown in
Tables~\ref{tab:top-src-ases} and~\ref{tab:top-dst-ases} (highlighted cells
discussed in detail).

\begin{table*}[!t]
    \caption{\emph{Top 10 source ASes (w.r.t. traffic volume by bytes) in April
    2019 (left) and 2020 (right) for samplepoint-F.}}
    \label{tab:top-src-ases}
    \resizebox{\linewidth}{!}{%

    \begin{tabular}{@{}lllrlllr@{}}
      \toprule
      \multicolumn{2}{c}{\textbf{\begin{tabular}[c]{@{}c@{}}Top 10 Source ASes\\ (April 2019)\end{tabular}}} &
        \multicolumn{1}{c}{\textbf{\begin{tabular}[c]{@{}c@{}}AS\\ Type\end{tabular}}} &
        \multicolumn{1}{c}{\textbf{\begin{tabular}[c]{@{}c@{}}Tfc. Vol.\\ {[}GiB{]}\end{tabular}}} &
        \multicolumn{2}{c}{\textbf{\begin{tabular}[c]{@{}c@{}}Top 10 Source ASes\\ (April 2020)\end{tabular}}} &
        \multicolumn{1}{c}{\textbf{\begin{tabular}[c]{@{}c@{}}AS\\ Type\end{tabular}}} &
        \multicolumn{1}{c}{\textbf{\begin{tabular}[c]{@{}c@{}}Tfc. Vol.\\ {[}GiB{]}\end{tabular}}} \\ \midrule
      AS15169 & \hr{GOOGLE}               & \hr{C}   & \hr{119.25} & AS2500  & WIDE-BB                          & T/A & 132.77 \\
      AS2500  & WIDE-BB              & T/A & 70.87  & AS19679 & \hg{DROPBOX}                          & \hg{C}   & \hg{24.31}  \\
      AS20940 & AKAMAI-ASN1          & C   & 67.72  & AS6185  & APPLE-AUSTIN                     & C   & 17.05  \\
      AS6185  & APPLE-AUSTIN         & C   & 50.16  & AS20940 & AKAMAI-ASN1                      & C   & 16.36  \\
      AS16509 & AMAZON-02            & E   & 37.21  & AS16509 & AMAZON-02                        & E   & 12.62  \\
      AS15133 & EDGECAST             & C   & 26.23  & AS17676 & GIGAINFRA Softbank               & T/A & 9.20   \\
      AS32934 & FACEBOOK             & C   & 25.82  & AS63783 & Chiba Univ. of Commerce          & T/A & 6.59   \\
      AS19679 & \hg{DROPBOX}              & \hg{C}   & \hg{19.20}  & AS13335 & CLOUDFLARENET                    & C   & 6.13   \\
      AS13335 & CLOUDFLARENET        & C   & 19.05  & AS16625 & AKAMAI-AS                        & C   & 5.61   \\
      AS2906  & NETFLIX AS-SSI       & C   & 9.49   & AS714   & APPLE-ENGINEERING                & C   & 5.39   \\ \bottomrule
    \end{tabular}
    }
\end{table*}

\subsubsection{\textbf{Source ASes}} We observe that for source ASes (see
Table~\ref{tab:top-src-ases}), the composition of ASes remains roughly the same:
The majority of incoming traffic volume is received from popular \acp{CDN} and
cloud providers, i.e., Akamai, Apple, Amazon, Dropbox, and Cloudflare. However,
while we also observe larger amounts of traffic to be received from other
prominent services and \acp{CDN} such as Google, Edgecast, Facebook, or Netflix
in 2019, these ASes disappear from the top 10 in 2020. We suspect that this is
due to university staff and students being moved off-campus, resulting in a
substantial decrease of popular end user content served from Facebook (AS32934)
and Netflix (AS2906), as they drop from 25.8~GiB (ranked 7\textsuperscript{th})
and 9.49~GiB (ranked 10\textsuperscript{th}) to less than 1~GiB of traffic
volume each (ranked 25\textsuperscript{th} and 26\textsuperscript{th}).

On the other hand, Google (AS15169) drops from 119.25~GiB (ranked
1\textsuperscript{st}) to 17.8~MiB (ranked 169\textsuperscript{th}), i.e., a
decrease by more than 99\%; similarly, Edgecast (AS15133) drops from 26.23~GiB
(ranked 6\textsuperscript{th}) to 6.15~MiB (ranked 291\textsuperscript{th}).
Therefore, these substantial decreases are likely due to changes in the AS
topology, rather than due to the pandemic.

However, although the overall traffic volume has decreased from 2019 to 2020 as
shown in Table~\ref{tab:apr-19vs20}, we find more traffic incoming from within
the WIDE network (AS2500), increasing from 70.87~GiB to 132.77~GiB, as well as
from Dropbox (AS19679), increasing from 19.2~GiB to 24.31~GiB. Considering the
COVID-19 restrictions, the latter observation indicates that clients in the WIDE
network are increasingly pulling data from their Dropbox storage, likely to
synchronize with work done remotely; note that this might also result in policy
violations depending on the classification of information, e.g., if sensitive
datasets are stored on such third-party servers.

\begin{table*}[!t]
  \caption{\emph{Top 10 destination ASes (w.r.t. traffic volume by bytes) in April
  2019 (left) and 2020 (right) for samplepoint-F.}}
    \label{tab:top-dst-ases}
    \resizebox{\linewidth}{!}{%

    \begin{tabular}{@{}lllrlllr@{}}
      \toprule
      \multicolumn{2}{c}{\textbf{\begin{tabular}[c]{@{}c@{}}Top 10 Destination ASes\\ (April 2019)\end{tabular}}} &
        \multicolumn{1}{c}{\textbf{\begin{tabular}[c]{@{}c@{}}AS\\ Type\end{tabular}}} &
        \multicolumn{1}{c}{\textbf{\begin{tabular}[c]{@{}c@{}}Tfc. Vol.\\ {[}GiB{]}\end{tabular}}} &
        \multicolumn{2}{c}{\textbf{\begin{tabular}[c]{@{}c@{}}Top 10 Destination ASes\\ (April 2020)\end{tabular}}} &
        \multicolumn{1}{c}{\textbf{\begin{tabular}[c]{@{}c@{}}AS\\ Type\end{tabular}}} &
        \multicolumn{1}{c}{\textbf{\begin{tabular}[c]{@{}c@{}}Tfc. Vol.\\ {[}GiB{]}\end{tabular}}} \\ \midrule
      AS2500  & WIDE-BB                      & T/A & 438.64 & AS2500  & WIDE-BB                        & T/A & 135.72 \\
      AS23799 & National Defense Academy     & T/A & 38.13  & AS17676 & GIGAINFRA Softbank             & T/A & 56.68  \\
      AS17676 & GIGAINFRA Softbank           & T/A & 15.42  & AS4837  & CHINA169-BACKBONE              & T/A & 26.21  \\
      AS715   & WOODYNET-2                   & C   & 12.72  & AS8068  & MICROSOFT-CORP-MSN             & C   & 11.61  \\
      AS7922  & COMCAST-7922                 & T/A & 5.29   & AS715   & WOODYNET-2                     & C   & 10.68  \\
      AS64238 & ASN-OARC                     & T/A & 4.12   & AS7922  & COMCAST-7922                   & T/A & 5.18   \\
      AS8075  & MICROSOFT-CORP-MSN           & C   & 2.21   & AS4717  & WIDE-BB                        & T/A & 4.67   \\
      AS2510  & INFOWEB FUJITSU              & T/A & 1.73   & AS8075  & MICROSOFT-CORP-MSN             & C   & 3.55   \\
      AS2518  & BIGLOBE                      & T/A & 1.47   & AS16276 & OVH                            & C   & 2.61   \\
      AS4725  & ODN SoftBank Corp.           & T/A & 1.44   & AS13445 & \hg{CISCO Webex LLC}                & \hg{C}   & \hg{2.22}   \\ \bottomrule
    \end{tabular}
    }
\end{table*}

\subsubsection{\textbf{Destination ASes}} In terms of outgoing traffic volume
(see Table~\ref{tab:top-dst-ases}), traffic is mostly destined toward
transit/access ASes in both April 2019 and 2020, although the set of destination
ASes partially changes between the years. In both years, most traffic is sent to
destinations within the WIDE network (AS2500 and AS4717), although the volume
has decreased significantly from 2019 (438.64~GiB) to 2020 (140.39~GiB). Traffic
destined to Dropbox (AS19679) only amounts to roughly 250--280~MiB in both April
2019 and 2020, indicating asymmetric synchronization behavior when compared to
the traffic volume originating from its AS. While most of the top 10 ASes do not
refer to specific services, we observe that Cisco Webex (AS13445) appears among
the top 10 destination ASes in 2020 (2.22~GiB) after increasing by a factor of
roughly ten, up from rank 47 in 2019 (210.39~MiB). The COVID-19 pandemic has
resulted in a large increase regarding video conferencing
traffic~\cite{Feldmann2020}; consequently, the increase of traffic volume toward
Cisco Webex AS in the WIDE network is very likely a result of the switch to
remote work.

\begin{summary}
    \emph{Takeaway:} 
    Comparing MAWI traffic data from April 2019 to 2020, we find that COVID-19
    prevention measures are reflected in the traffic mix, as protocols such as
    OpenVPN and \texttt{rsync} see significant increases in terms of traffic
    volume over both address families (up to 3082.7\%). Weekday-weekend patterns
    have also become less pronounced. In terms of source and destination ASes,
    we find that traffic volume received from \acp{CDN} has reduced, while
    traffic from Dropbox (cloud storage) and traffic to Cisco Webex (video
    conferencing) have increased, which indicates that users move off-campus and
    shift toward remote work due to COVID-19 restrictions.
\end{summary}

\section{Impact of Evolving Protocols}\label{sec:longitudinal}

\begin{figure*}[!t]
  \centering
  \begin{subfigure}{.5\textwidth}
    \centering
    \includegraphics[width=\linewidth]{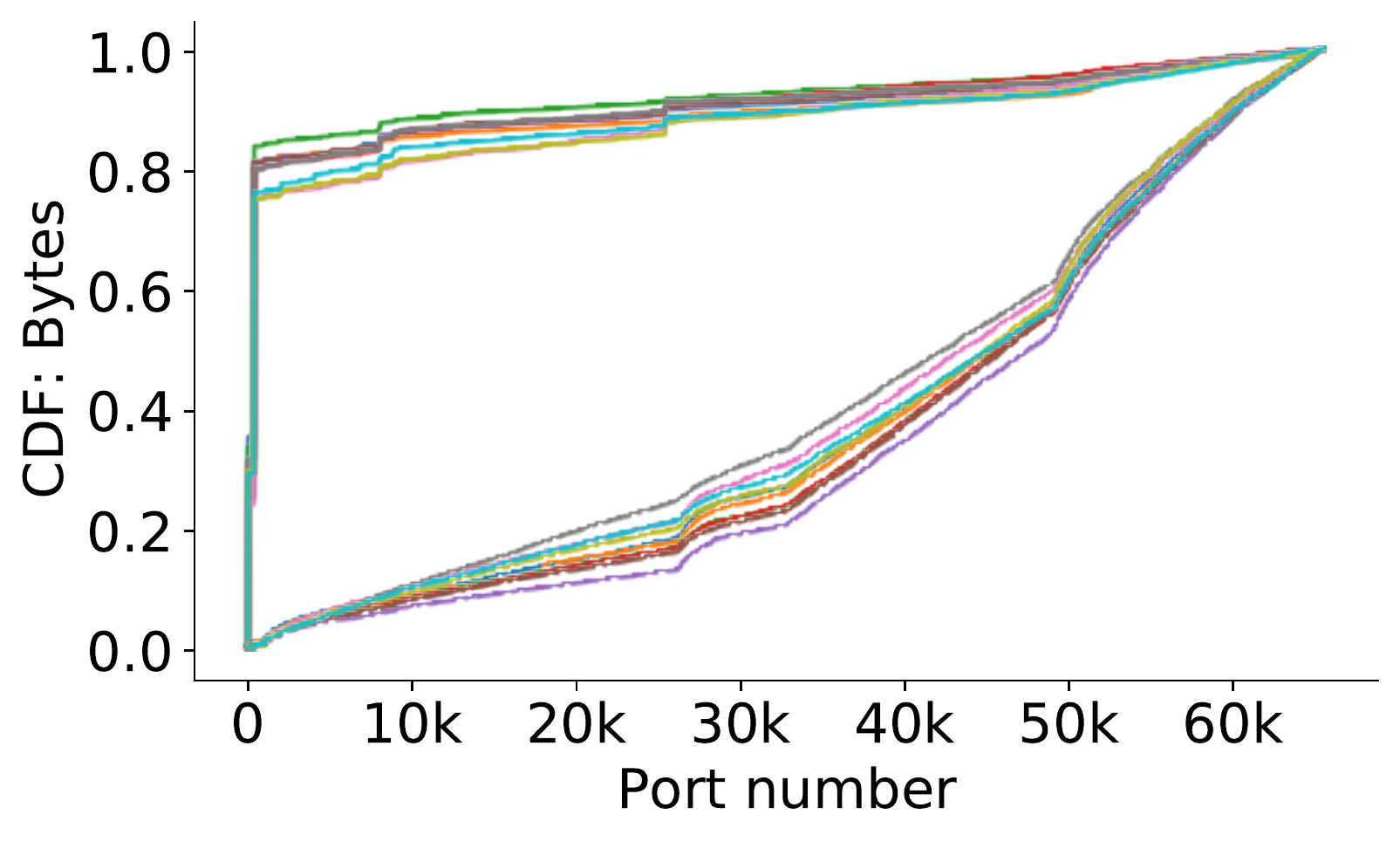}
    \caption{\em CAIDA SPNY.}
    \label{fig:ports_4A}
  \end{subfigure}\hfill
  \hfill
  \begin{subfigure}{.5\textwidth}
    \centering
    \includegraphics[width=\linewidth]{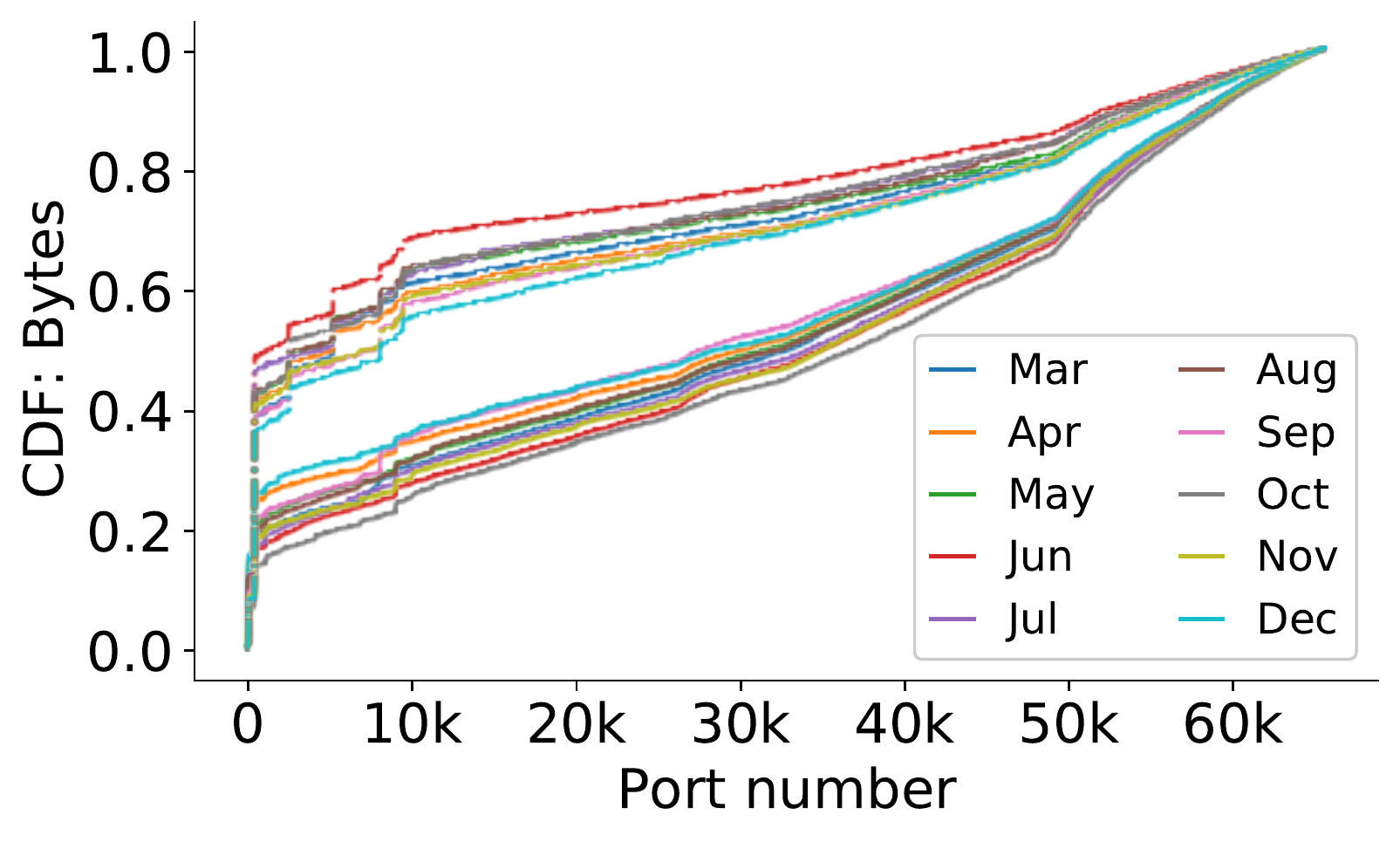}
    \caption{\em CAIDA NYSP.}
  \label{fig:ports_4B}
  \end{subfigure}
  \caption{\em IPv4 traffic distribution over TCP/UDP ports in bytes for CAIDA,
  separated by SPNY (a) and NYSP (b). Destination ports are represented
  by dashed lines, source ports by solid ones. 80\% of source ports on SPNY are in
  the well-known port range, indicating a major portion of server-to-client
  traffic from S\~{a}o Paulo to New York.}
  \label{fig:ports_4}
  \end{figure*}

As shown in the previous section, the Internet traffic composition can change
drastically depending on recent, life-changing events. However, the Internet
also evolves over time naturally, resulting in a shift regarding protocols and
services used, which we will focus on in this section.

\subsection{IPv4 and IPv6 Traffic Analysis}\label{sub:analysis-IP}

\subsubsection{\textbf{IPv4 Traffic}}
Fig.~\ref{fig:ports_4A} shows the traffic distribution over the source and
destination ports for IPv4 on SPNY. Around 80\% of the traffic originates from
source ports of the well-known port range (1--1024), which indicates that most
of the traffic on SPNY comes from servers running the associated services. On
the other hand, the destination ports have no such distinct behavior in the
well-known port range. Instead, we see a mostly linear trend, with three minor
spikes in the CDF: The first cluster is located at around port 28000, though
cause and purpose are not clear. The second spike starts from port 32768 onwards,
where the standard ephemeral port range for Linux starts. The third begins at
port 49152, which is the start of the ephemeral port range for Windows since
Windows Vista~\cite{Windows_ports}.

For NYSP, Fig.~\ref{fig:ports_4B} shows the traffic distribution for source and
destination ports. About 40\% of traffic ends in destination ports which are in
the well-known port range, which is expected as it represents the respective
client traffic to the opposite direction (SPNY). In addition, we see around 80\%
of traffic coming from ports above 1024.  This leaves 20\% that come from ports
below 1024, indicating that there is a moderate share of server-to-client
traffic on NYSP as well. Again, the increase of the curve at the start of the
Windows ephemeral port range can be observed, this time for both source and
destination ports.  The destination ports curve further exhibits a step-wise
pattern up to port 10000, which we observe due to popular applications within
the registered ports range.

\begin{figure}[!t]
  \centering
  \includegraphics[width=\linewidth]{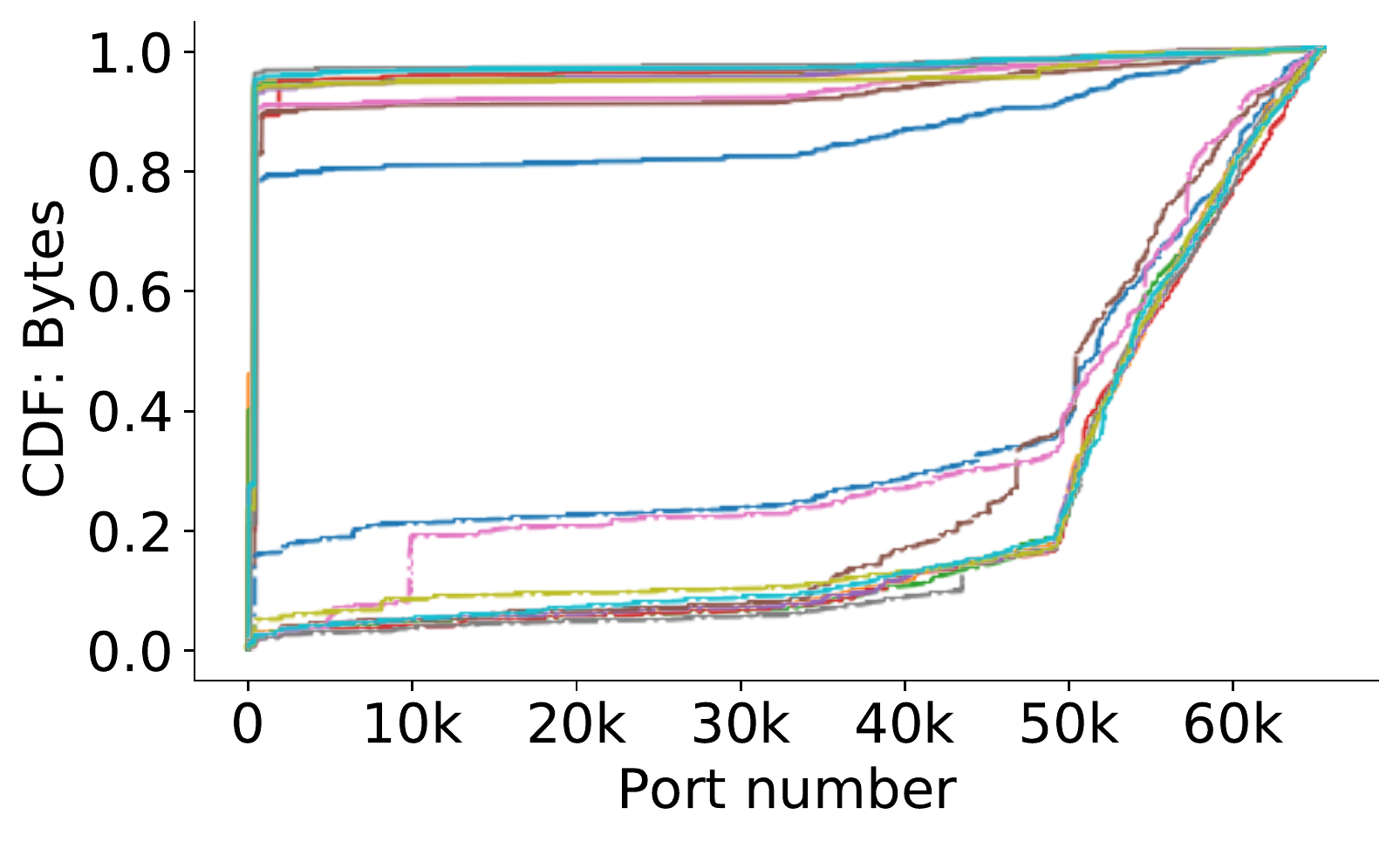}
  \caption{\em IPv4 traffic distribution over TCP/UDP ports in bytes for MAWI.
  Destination ports are represented
  by dashed lines, source ports by solid ones. The port distribution is similar to the distribution observed for SPNY.}
\label{fig:ports_4MAWI}
\end{figure}
Fig.~\ref{fig:ports_4MAWI} shows the distribution for the MAWI link and exhibits
a similar distribution as SPNY. This indicates that it mostly carries
server-to-client traffic over the upstream link.

\subsubsection{\textbf{IPv6 Traffic}} 
\begin{figure}[!t]
  \centering
  \includegraphics[width=\linewidth]{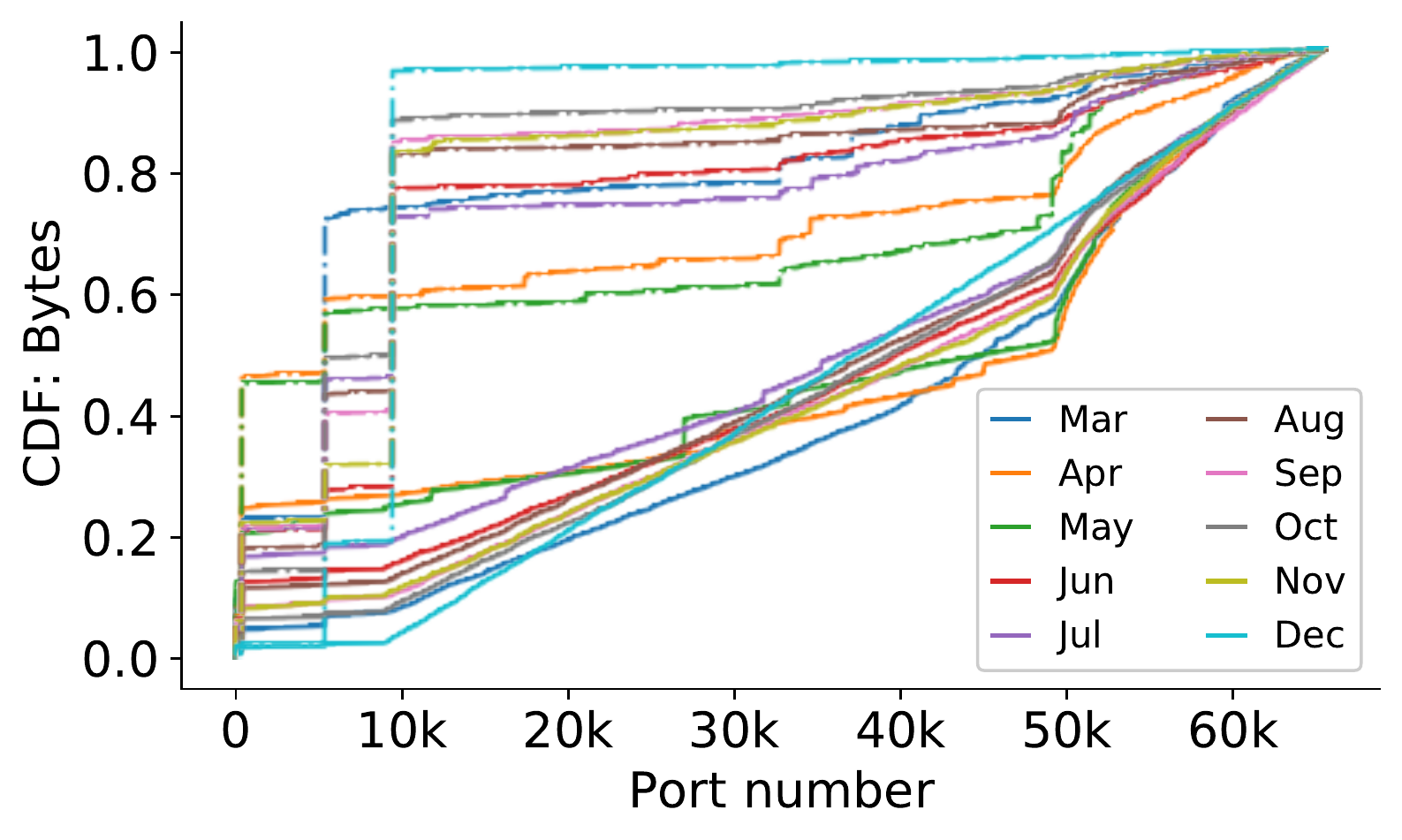}

\caption{\em IPv6 traffic distribution over TCP/UDP ports in bytes for
  direction NYSP. Destination ports are represented by dashed lines, source
  ports by solid ones. Source ports are similarly distributed as over IPv4,
  however, destination ports exhibit spikes at ports 5443 and 9501.
}
\label{fig:ports_6B}
\end{figure}

\begin{figure}[!t]
  \centering
 \includegraphics[width=\linewidth]{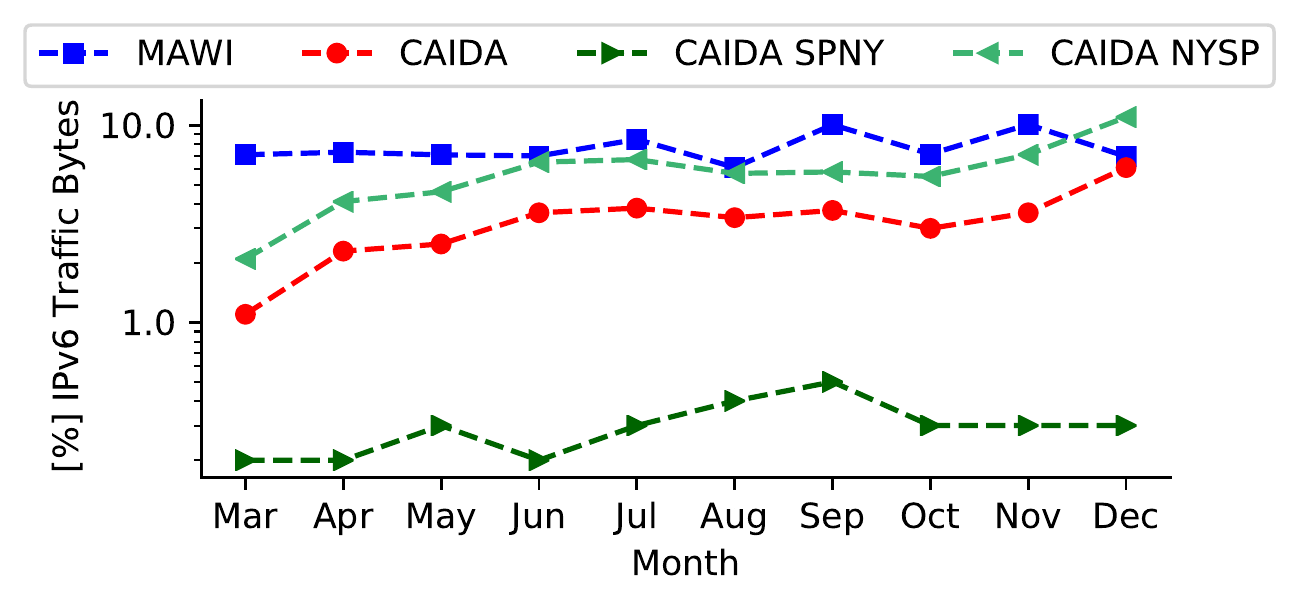}
 \caption{\em Distribution of IPv6 traffic share for CAIDA (SPNY, NYSP, and in
 total) and MAWI. IPv6 traffic increases from 1.1\% to 6.1\% for CAIDA and is consistently at around 9\% for MAWI. 
 }
 \label{fig:ip_prot}
\end{figure}

Fig.~\ref{fig:ports_6B} visualizes the
traffic distribution by ports in NYSP over IPv6. We do not include the graphs
for IPv6 on SPNY here, as IPv6 traffic on SPNY only accounts for 0.5\% of
overall traffic at its peak (see Fig.~\ref{fig:ip_prot}). For MAWI, the
distribution of ports over IPv6 is similar to its IPv4 distribution, thus also
not shown. The distribution of NYSP over IPv6, however, shows an interesting
pattern when compared to its IPv4 counterpart:
Around 20\% of the source ports are in the well-known port range; traffic shares
increase linearly from port 1024 onward, with a visible spike after port 49152.
Regarding destination ports, we observe significant spikes on the same two ports
inside the registered port range every month, namely ports 5443 and 9501;
traffic on these ports almost has the same source and destination IP prefixes.

Fig.~\ref{fig:ip_prot} shows the distribution of IPv6 traffic for CAIDA (SPNY,
NYSP, and total) and MAWI. IPv4 is by far the more popular address family; the
IPv6 traffic share on the CAIDA traces is just 1.1\% in March 2018 but
eventually grows to 6.1\% despite some fluctuations at the end of the
measurements in December 2018. Traffic coming from S\~{a}o Paulo (SPNY) barely
carries any IPv6 packets; with values between 0.1\% and 0.5\%, this is even
lower than observed worldwide years ago in
2013~\cite{DBLP:conf/pam/RichterCSFW15}. Nevertheless, we notice that
destination port 5443 accounts for a large traffic share in this direction. We
suspect that the traffic is related to DNS services, as the flows originate from
port 53, and as port 5443 is further used for DNSCrypt services.
In the other direction, traffic coming from New York (NYSP) reaches up to 11\%
of IPv6 traffic share.
CAIDA offers insights on the packet sizes for IPv6~\cite{caida_trace_stats},
from which we can see that IPv6 traffic on NYSP has a very low median packet
size between 70 and 100 bytes but a much bigger mean packet size that is 3--5
times higher. For IPv4, we see the opposite (not shown in the figure), where the
mean packet size is almost half the median packet size. Therefore, we assume
that the applications using ports 5443 and 9501 carry few but big packets and
almost solely push the IPv6 traffic share (defined by the number of transferred
bytes) on NYSP. This large volume of intra-AS traffic is used to synchronize
data between content replicas across data centers (further analysis withheld to preserve the
anonymity of the trace); note that IPv6 is known to be heavily used for
content delivery purposes~\cite{DBLP:conf/sigcomm/CzyzAZIOB14}.
On the other hand, the MAWI traces already start off with 7.1\% IPv6 traffic,
but do not show a growing trend over time. Although some months have over 10\%
of IPv6 traffic, the percentage is even slightly less in December than at the
beginning of March. When compared to up to date IPv6 measurements mentioned in
\textsection\ref{sec:related_work}, both datasets show much less IPv6 adoption
overall. As explained in previous work~\cite{DBLP:conf/sigcomm/CzyzAZIOB14},
quantifying IPv6 adoption relies heavily on the metric used, as well as the
vantage point location: Google~\cite{google_ipv6} as well as APNIC~\cite{apnic}
use different metrics, e.g., counting the number of users accessing their
service via IPv6 in Google's case. The same work measures 0.68\% of traffic over
IPv6 in 2014. When we compare this to the most recent percentages on the CAIDA
dataset, this is an approximate 800\% increase in four years. When compared to
MAWI's 6.9\% in December 2018, this even represents a growth of more than 900\%.
In contrast, Czyz \emph{et al.}~\cite{DBLP:conf/sigcomm/CzyzAZIOB14} observed a
yearly 400\% increase in IPv6 traffic between 2012 and 2014.

\subsection{Transport Layer Traffic Analysis}
\label{sec:transport_prot}

For the different transport protocols, we determine the traffic volume shares in
terms of bytes within the MAWI dataset (CAIDA omitted for brevity).
Since FB-Zero runs over TCP and uses the same port as HTTPS, we cannot
distinguish FB-Zero from TCP in our traces. As expected, TCP traffic dominates
with more than 90\% traffic share. This is followed by QUIC with an average of
7\% traffic share, which is a very high share for being a relatively new
protocol; this observation is in-line with prior
work~\cite{DBLP:conf/conext/TrevisanGDMM18, Sandvine:report,
DBLP:conf/sigcomm/LangleyRWVKZYKS17}. UDP (besides QUIC) has a depleting traffic
share of around 1.5\%. However, we do not observe much \ac{MPTCP} traffic in our
analysis ($\approx$1\%):
No clear longitudinal trend is visible in the results across the months; the
traffic share of MPTCP remains similar with slight variations, which can likely
be attributed to the short (15 minutes) time frame of the traces.

\begin{summary}
  \emph{Takeaway:} Overall, we find that the CAIDA dataset has an average of 4\%
  IPv6 traffic share, consistently lower to that of the MAWI dataset
  ($\sim$9\%). We further notice the CAIDA link to carry different traffic
  depending on the observed direction: SPNY carries primarily server-to-client
  traffic, whereas NYSP carries client-to-server traffic instead. Regarding
  transport protocols, TCP to accounts for 91.5\% of the traffic and UDP for
  1.5\%. As an evolving protocol, the traffic share of QUIC is already seen to
  be at 7\%.
\end{summary}

\subsection{Web Browsing Traffic Analysis}
\label{sec:Web_dominance}

Fig.~\ref{fig:apps} presents the top applications from the CAIDA (top) and MAWI
(bottom) datasets over IPv4 and IPv6 in terms of traffic shares. Classified
applications other than the ones displayed in the legend are grouped into the
category ``misc''; applications that could not be mapped are subsumed under
``unclassified'' instead. Note that XMPP is only popular over IPv4, while NTP
(CAIDA) and \texttt{rsync} (MAWI) are only popular over IPv6, which is why they
are only shown for one address family.

\subsubsection{\textbf{IPv4 Application Mix}} 
\begin{figure}[!t]
  \centering
  \includegraphics[width=\linewidth]{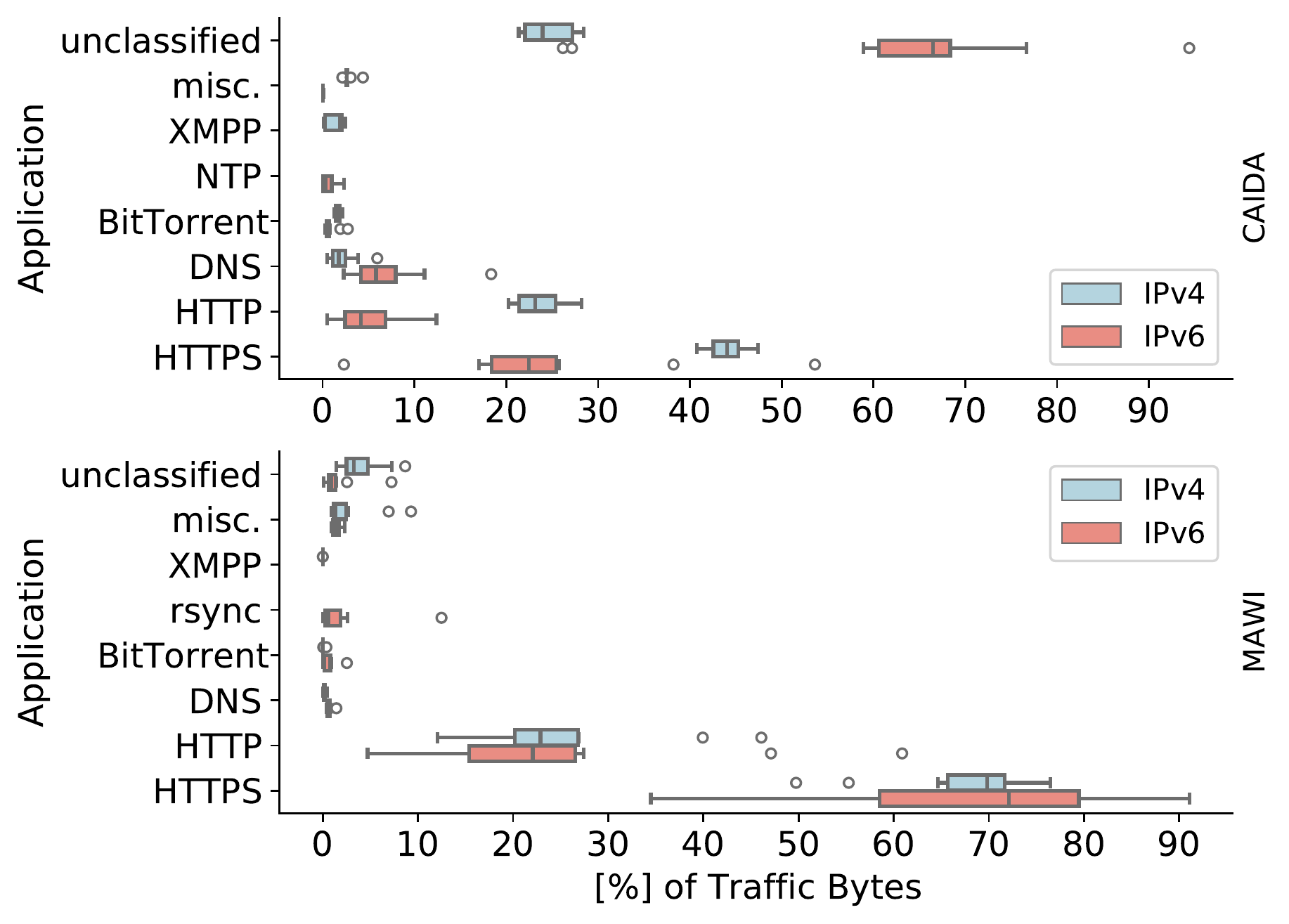}
  \caption{\em Boxplots over traffic shares by month for most popular
  applications for CAIDA (top) and MAWI (bottom) over both IPv4 (blue) and IPv6
  (red). \textbf{IPv4 --} CAIDA data over IPv4 consists of 60\% to 70\% HTTP(S)
  with a 2-to-1 HTTPS to HTTP ratio. For MAWI, the Web makes up more than 90\%
  of all IPv4 traffic with a 3-to-1 HTTPS to HTTP ratio. \textbf{IPv6 --} The
  IPv6 mix appears similar to the IPv4 one for MAWI; however, CAIDA has mostly
  unclassified applications (on ports 5443 and 9501).}
  \label{fig:apps}
\end{figure}

In this section, we look into the
applications observed over IPv4 (see Fig.~\ref{fig:apps}, blue boxes). 
The CAIDA IPv4 mix primarily consists of HTTPS, followed by HTTP, both together
accounting for 60--70\% of traffic. 
Note that encrypted Web traffic such as HTTP/2 and FB-Zero is subsumed under
HTTPS in our results. When compared to the 2013 application mix from Richter
\emph{et al.}~\cite{DBLP:conf/pam/RichterCSFW15}, where HTTPS contributes less
than 10\% to the overall traffic, it is clear that HTTPS has been growing at the
expense of HTTP. Nevertheless, HTTPS and HTTP together still account for a
similar fraction of all traffic. In the previous study~\cite{DBLP:conf/pam/RichterCSFW15}, the ratio between HTTPS to
HTTP was 1-to-6 in 2013, however, we see an approximately 2-to-1 HTTPS to HTTP
ratio in the current data. More recent work also measured a 3-to-1 HTTPS to HTTP
ratio by the end of 2017~\cite{DBLP:conf/conext/TrevisanGDMM18}.

The MAWI data presents a different view:
On average, the HTTPS to HTTP ratio is around 3-to-1, which is higher than in
the CAIDA traces. Previous research on the same MAWI monitor showed a 2-to-3
HTTPS to HTTP ratio in 2017~\cite{DBLP:conf/uss/FeltBKPBT17}, which means that
the MAWI traffic is trending toward more encrypted Web traffic. HTTP and HTTPS
are also accountable for a much larger fraction of traffic compared to the CAIDA
data, with more than 90\% of traffic share in some months. This dominance of
Web-related traffic can be clearly seen in the MAWI traces, whereas the CAIDA
traces exhibit a more diverse application mix.

We further inspect the application mix for both directions SPNY and NYSP of the
CAIDA monitor separately by month. We observe that traffic on SPNY carries more
HTTPS and HTTP than its counterpart: Web traffic accounts for consistently more
than 70\% on SPNY but always less than 60\% on NYSP, in December 2018 even less
than 50\%. The discrepancy for HTTP in particular is striking here: Direction
SPNY is composed of around 20\%-30\% of HTTP traffic, whereas there is only
10\%-20\% on NYSP. 
NYSP traffic sees a relatively large fraction of XMPP, BitTorrent, DNS, and
other classified applications instead, which indicates asymmetry between the two
directions regarding the application mix.

\subsubsection{\textbf{IPv6 Application Mix}}
The application mixes over IPv6 (see Fig.~\ref{fig:apps}, red boxes) exhibit a
different behavior. With respect to the CAIDA dataset and
Fig.~\ref{fig:ip_prot}, we see that the majority of IPv6 traffic is carried on
NYSP. As such, CAIDA's IPv6 application mix is mainly determined by traffic in
that direction. We notice more than 50\% of unclassified traffic across all
months (except for April and May 2018), which comes from two applications on
ports 5443 and 9501, following the aforementioned reasoning. This part of the
traffic also increases over time, to a point where it is over 90\% of IPv6
traffic in December 2018. The rest is mainly HTTP(S) and DNS, with a much higher
HTTPS to HTTP ratio than seen over IPv4. The largest part of the DNS traffic on
IPv6 comes from traffic in SPNY, as more than half of SPNY's IPv6 traffic is
DNS. Earlier measurements show similar behavior to our observations for SPNY,
where IPv6 is mainly used for DNS queries (85\% DNS traffic on average in
2009~\cite{DBLP:conf/pam/KarpilovskyGPRS09}). The overall picture of the IPv6
traffic in the CAIDA dataset is contrary to the trend observed by Czyz \emph{et
al.}~\cite{DBLP:conf/sigcomm/CzyzAZIOB14}, who suggest that the IPv6 application
mix is adapting to the one of IPv4.

When we look at the MAWI IPv6 mix in Fig.~\ref{fig:apps}, HTTPS and HTTP
together account for more than 90\% of traffic in the median case over the
months, which is even a larger share than observed in the IPv4 mix. Furthermore,
DNS accounts for a very small part of the traffic only, similar to IPv4. The
results we see head more toward the direction of Czyz \emph{et al.}'s
observations~\cite{DBLP:conf/sigcomm/CzyzAZIOB14}.
Note that the plot displays \texttt{rsync} instead of NTP for CAIDA but shows
rsync for MAWI over IPv6 in Fig.~\ref{fig:apps}, as \texttt{rsync} does not
appear at all in the CAIDA IPv6 mix, whereas it is popular in the MAWI traces.
At its peak in July 2018, \texttt{rsync} even reaches 12.5\% of IPv6 traffic
volume.

\begin{summary}
  \emph{Takeaway:} In terms of Web traffic, we find the HTTPS to HTTP ratio
  growing overall, ranging from 2-to-1 up to 3-to-1. Evolving transport
  protocols, such as QUIC and FB-Zero, increasingly contribute to Web traffic at
  the expense of TCP, which further increases the HTTPS traffic share. Moreover,
  comparing the MAWI and CAIDA data (including both directions), highlights how
  different the traffic composition can be depending on the selected samples.
\end{summary}

\section{Limitations and Future Directions}\label{sec:limitations}
We are aware that traffic varies for different hours and
weekdays~\cite{DBLP:conf/conext/TrevisanGDMM18}, and that traffic captures from
both CAIDA and MAWI links are different in length.
While similar results have been already observed by previous work for the same
links, our study updates, combines, and continues these studies by putting the
monitored traffic in contrast with each other. However, we also acknowledge that
different types of traffic are carried by the monitored links. Thus, the
presented results are specific to these links and cannot be generalized.
Simultaneously, finding differing observations due to inherent differences of
vantage points is also a point we aim to highlight with this study. 

Future studies should further look into the different types of end users. While
the MAWI datasets largely cover users from an educational/research network, the
constellation of end users is unknown and cannot be determined for the CAIDA
data due to header anonymization.

Moreover, the heuristic used for identification is limited, as it can only
identify well-known protocols and applications. We acknowledge that a port-based
identification will underestimate the observed traffic shares, however, an
exhaustive application identification method is not the focus of this work. Due
to the datasets anonymizing IP prefixes and addresses at the cost of the report
transparency~\cite{FragkouliAF19}, we cannot incorporate additional information
of the traces, such as sources and destinations, over a more extended period of
time for a more fine-grained identification. As such, this does not allow us to
draw conclusions for the endpoints of the measured traces, e.g., regarding
potential geographic relationships. The analysis of source and destination ASes
is only possible for \textsection~\ref{sub:covid-ases} due to the support from
MAWI, who provided us with non-anonymized data. The AS-based analysis can,
therefore, not be extended to the whole monitoring period of the MAWI monitor.
We are aware of other evolving protocols and protocol extensions/updates, such
as HTTP/2~\cite{DBLP:journals/rfc/rfc7540}, TLS
1.3~\cite{DBLP:journals/rfc/rfc8446}, DNS over TLS
(DoT)~\cite{DBLP:journals/rfc/rfc7858,DBLP:journals/rfc/rfc8310}, or DNS over
HTTPS (DoH)~\cite{DBLP:journals/rfc/rfc8484}. However, traffic using HTTP/2, TLS
1.3, or DoH cannot be accurately identified based on the public trace data, as
the monitors drop all packet information above the transport layer. In addition
to this, the mentioned protocols do not yet contribute meaningfully to Internet
traffic shares, as we find DoT traffic shares to be negligible (< 0.1\%), for
instance.

Regarding future directions, extending the analysis by considering a vantage point in Europe would add further insights. However, a similar dataset is not available in the public domain due to GDPR regulations. Further, the datasets that are indeed online are a bit dated. To this end, operators can consider making more recent datasets available in the public domain to allow researchers better provide insights into the trends of Internet traffic composition.

Additionally, an online dashboard to visualize and present the trends in real-time would be very valuable for planning of future short-term events (similar to COVID-19) that dramatically led to change in Internet usage behavior. Operators reactively adapted to such changes but an online dashboard can help visualize such changes more quickly, leading to a more proactive approach to network capacity planning for future events.

\section{Conclusion}\label{sec:conclusion}

Evaluating the impact of the COVID-19 lockdown, we observed that the overall
traffic volume decreased on the MAWI link over both address families, although
we saw a surge in traffic volume from Dropbox and toward Cisco Webex ASe; we
similarly observed a substantial relative increase in traffic volume for
applications (OpenVPN and \texttt{rsync}) that facilitate remote work. More
surprisingly, we found that although the application mix over IPv4 remained
similar, it changed substantially over IPv6 during the lock-down in favor of
HTTP over HTTPS.

We further presented a longitudinal view of the Internet by analyzing two trace
datasets from Brazil/USA (CAIDA) and Japan (MAWI).
The traffic distribution to the different source and destination
ports on the CAIDA traces revealed that most of the traffic in these traces
come from servers in S\~{a}o Paulo and clients in New York. SPNY carries more Web traffic on the application layer, but NYSP has a higher HTTPS to HTTP ratio.
We saw that IPv6 usage is increasing on CAIDA's traces and is increasingly
stable for MAWI's, with peak values of 10.1\% IPv6 traffic share in 2018 (12.2\%
in 2020). Regarding the transport layer, we find that TCP dominates the
transport layer traffic with 90\% of the traffic. QUIC achieves 7\%, despite
being a relatively new protocol. For applications, we discovered that the
traffic mix for IPv4 consists mostly of Web traffic in both datasets, with a
higher and still growing part of HTTPS. In the CAIDA data, the HTTPS to HTTP
ratio is 2-to-1, whereas the ratio is 3-to-1 in the MAWI dataset. Earlier traces
on the same MAWI monitor from two years ago showed different results with HTTP
in the lead~\cite{DBLP:conf/uss/FeltBKPBT17}. In the MAWI traces, HTTP and HTTPS
also dominate the application mix with more than 90\% of traffic in some months,
indicating the high share of Web traffic.
We saw a vastly different application mix for the CAIDA dataset over IPv6
when compared to its IPv4 counterpart, with the majority of traffic utilized
to synchronize content replicas.
In the MAWI data, we instead notice that the shape of IPv6 traffic is more
similar to that of IPv4 traffic.

In conclusion, our analysis provided a relative comparison of evolving protocols
from two very different vantage points across the globe during the same time
period. We demonstrated that the evolution trends can vary based on where the
observations are made, despite applying the same methods and analyses to the
datasets. This further presents a novel perspective on related work that largely
reported trends collected from datasets in one particular region, such as Europe
or North America. Future work, particularly on evolving protocols and short-term
sudden effects, should therefore take such aspects into consideration.

\section*{Reproducibility Considerations} The CAIDA~\cite{CAIDA:monitor} and the
MAWI~\cite{MAWI} datasets analyzed in \textsection\ref{sub:covid-mix} and
\textsection\ref{sec:longitudinal} are publicly available. The scripts and
Jupyter notebooks used for processing, aggregating, and analyzing the data will
be released on Github to facilitate reproducibility of results.

{
\balance{
  \bibliographystyle{IEEEtran}
  \bibliography{bibtex/bib/IEEEabrv.bib,bibtex/bib/IEEEexample.bib}
  }
}
\end{document}